%% file: paper_prl.tex
\documentclass[twocolumn,showpacs,aps,prl,superscriptaddress,letterpaper]{revtex4}

\usepackage{graphicx}
\usepackage{dcolumn}
\usepackage{amsmath}
\usepackage{epsfig}
\usepackage{multirow}

\long\def\inst#1{\par\nobreak\kern 4pt\nobreak
    {\itshape #1}\par\vskip 10pt plus 3pt minus 3pt}
\RequirePackage{xspace}
\usepackage{relsize}

\def\Bztorhozrhoz {\ensuremath{\Bz \to \rho^0 \rho^0 }\xspace}
\def\Btozz {\ensuremath{\Bz \to \rho^0 \rho^0 }\xspace}

\def\Bztorhoprhom {\ensuremath{\Bz \to \rho^+ \rho^- }\xspace}
\def\Bptorhozrrhop {\ensuremath{\Bp \to \rho^+ \rho^0 }\xspace}

\def\babar{\mbox{\slshape B\kern-0.1em{\smaller A}\kern-0.1em
    B\kern-0.1em{\smaller A\kern-0.2em R}}}
\def\Bbar    {\kern 0.18em\overline{\kern -0.18em B}{}\xspace}
\def\Dbar    {\kern 0.18em\overline{\kern -0.18em D}{}\xspace}
\def\Kbar    {\kern 0.18em\overline{\kern -0.18em K}{}\xspace}
\def\pep2{PEP-II}
\mathchardef\Upsilon="7107
\newcommand{\optbar}[1]{\shortstack{{\tiny (\rule[.4ex]{1em}{.1mm})}
  \\ [-.7ex] $#1$}}
\def\BorBbar    {\kern 0.18em\optbar{\kern -0.18em B}{}\xspace}
\def\DorDbar    {\kern 0.18em\optbar{\kern -0.18em D}{}\xspace}
\def\KorKbar    {\kern 0.18em\optbar{\kern -0.18em K}{}\xspace}

\def\babar{\mbox{\slshape B\kern-0.1em{\smaller A}\kern-0.1em
    B\kern-0.1em{\smaller A\kern-0.2em R}}}
\def\Dbar    {\kern 0.18em\overline{\kern -0.18em D}{}\xspace}
\def\B       {\ensuremath{B}\xspace}
\def\Bbar    {\kern 0.18em\overline{\kern -0.18em B}{}\xspace}

\def\BB      {\ensuremath{B\Bbar}\xspace}
\def\Bz      {\ensuremath{B^0}\xspace}
\def\Bzb     {\ensuremath{\Bbar^0}\xspace}
\def\BzBzb   {\ensuremath{\Bz {\kern -0.16em \Bzb}}\xspace}
\def\Bu      {\ensuremath{B^+}\xspace}
\def\Bub     {\ensuremath{B^-}\xspace}
\def\Bp      {\ensuremath{\Bu}\xspace}

\def\BpBm    {\ensuremath{\Bu {\kern -0.16em \Bub}}\xspace}

\def\CP                {\ensuremath{C\!P}\xspace}
\def\pep2{PEP-II}
\mathchardef\Upsilon="7107
\def\Y#1S{\ensuremath{\Upsilon{(#1S)}}\xspace}% no space before {...}!

\def\FourS {\Y4S}

\def\BR         {{\ensuremath{\cal B}\xspace}}
\def\Bztorhozrhoz {\ensuremath{\,\Bz \to \rho^0\rho^0}\xspace}
\def\Bztorhoprhom {\ensuremath{\,\Bz \to \rho^+\rho^-}\xspace}

\def\Bztorhozfz {\ensuremath{\,\Bz \to \rho^0 f_0}\xspace}
\def\Bztofzfz   {\ensuremath{\,\Bz \to f_0 f_0}\xspace}

\def\DeltaE {\ensuremath{\Delta E}\xspace}
\def\mes{\ensuremath{m_{\mathrm{ES}}}\xspace}

\newcommand{\tev}{\ensuremath{\mathrm{\,Te\kern -0.1em V}}\xspace}
\newcommand{\gev}{\ensuremath{\mathrm{\,Ge\kern -0.1em V}}\xspace}
\newcommand{\mev}{\ensuremath{\mathrm{\,Me\kern -0.1em V}}\xspace}
\newcommand{\kev}{\ensuremath{\mathrm{\,ke\kern -0.1em V}}\xspace}
\newcommand{\ev}{\ensuremath{\mathrm{\,e\kern -0.1em V}}\xspace}
\newcommand{\gevc}{\ensuremath{{\mathrm{\,Ge\kern -0.1em V\!/}c}}\xspace}
\newcommand{\mevc}{\ensuremath{{\mathrm{\,Me\kern -0.1em V\!/}c}}\xspace}
\newcommand{\gevcc}{\ensuremath{{\mathrm{\,Ge\kern -0.1em V\!/}c^2}}\xspace}
\newcommand{\mevcc}{\ensuremath{{\mathrm{\,Me\kern -0.1em V\!/}c^2}}\xspace}

\newcommand{\jprlBase}       {Phys.\ Rev.\ Lett.\xspace}
\newcommand{\jprl}      [1]  {\jprlBase\ {\bf #1}}

\def\u     {\ensuremath{u}\xspace}

\def\d     {\ensuremath{d}\xspace}

\def\b     {\ensuremath{b}\xspace}

\def\B       {\ensuremath{B}\xspace}
\def\mes        {\mbox{$m_{\rm ES}$}\xspace}
\def\DeltaE     {\mbox{$\Delta E$}\xspace}

\def\epem       {\ensuremath{e^+e^-}\xspace}

\newcommand{\ie}{{\em i.e.}}
\newcommand{\eg}{{\em e.g.}}

\newcommand{\BABARPubYear}     {08}
\newcommand{\BABARPubNumber}  {031}

\newcommand{\SLACPubNumber} {13329}

\begin{document}

\onecolumngrid
\begin{flushright}
\babar-PUB-\BABARPubYear/\BABARPubNumber \\
SLAC-PUB-\SLACPubNumber \\
%arXiv:\LANLNumber (hep-ex) \\
%% %March 2008 \\
\end{flushright}

\vspace{-\baselineskip}
%%\twocolumngrid
% Title of the paper
\title{
\large \bfseries \boldmath
Measurement of the Branching Fraction, Polarization, and \CP\
Asymmetries in \Btozz\ Decay, and Implications for the CKM Angle $\alpha$
}

%%%%%%%%%%%%%%%%%%%%%%%%%%%%%%%%%%%%%%%%%%%%%%%%%%%%%%%%%%%%%%%%%%%%%%%%%%%%%%%%%%
%% author list :
%
\input authors_jul2008
\date{July 30, 2008}

%%%%%%%%%%%%%%%%%%%%%%%%%%%%%%%%%%%%%%%%%%%%%%%%%%%%%%%%%%%%%%%%%%%%%%%%%%%%%%%%%%
%                             Abstract                                          %%
%%%%%%%%%%%%%%%%%%%%%%%%%%%%%%%%%%%%%%%%%%%%%%%%%%%%%%%%%%%%%%%%%%%%%%%%%%%%%%%%%%

\begin{abstract}
We study \Btozz\ decays in a sample of $465\times10^6$
$\Upsilon (4S)\rightarrow B\kern 0.18em\overline{\kern -0.18em B}$
events collected with the \babar\ detector at the
PEP-II asymmetric-energy $e^+e^-$ collider located at the Stanford
Linear Accelerator Center (SLAC). 
We measure the branching fraction 
$\BR = (0.92\pm 0.32\pm 0.14)\times 10^{-6}$
and longitudinal polarization fraction
$f_L = 0.75^{+0.11}_{-0.14}\pm 0.04$, where the first uncertainty is
statistical and the second is systematic.
The evidence for the \Btozz\ signal has a significance of $3.1$
standard deviations, including systematic uncertainties. 
We investigate the proper-time dependence of the 
longitudinal component in the decay and measure the
$C\!P$-violating coefficients 
$S^{00}_L = (0.3\pm0.7\pm0.2)$
and $C^{00}_L = (0.2\pm0.8\pm0.3)$.
We study the implication of these results for 
the unitarity triangle angle $\alpha$.
\end{abstract}

% PACS, the Physics and Astronomy Classification Scheme.

\maketitle
%%%%%%%%%%%%%%%%%%%%%%%%%%%%%%%%%%%%%%%%%%%%%%%%%%%%%%%%%%%%%%%%%%%%%%%%%
% INTRODUCTION
%%%%%%%%%%%%%%%%%%%%%%%%%%%%%%%%%%%%%%%%%%%%%%%%%%%%%%%%%%%%%%%%%%%%%%%%%

Measurements of \CP-violating asymmetries 
test the flavor structure of the standard model by 
over-constraining the Cabibbo-Kobayashi-Maskawa (CKM)
quark-mixing matrix~\cite{CabibboKobayashi}.
The time-dependent \CP asymmetry in the decays of \Bz\ or \Bzb\ mesons
to a \CP eigenstate dominated by the tree-level amplitude
$\b \to \u{\bar\u}\d$
measures $\sin 2\alpha_\mathrm{eff}$, where
$\alpha_\mathrm{eff}$ differs from the CKM unitarity
triangle angle $\alpha\equiv
\arg\left[-V_{td}^{}V_{tb}^{*}/V_{ud}^{}V_{ub}^{*}\right]$ by a
quantity  $\Delta\alpha$ accounting for the contributions from 
loop (penguin) amplitudes.
The value of $\Delta\alpha$ can be extracted from an analysis of
the full set of isospin-related channels~\cite{gronau90}.

Since the tree contribution to the $B^0\to\rho^0\rho^0$
~\cite{footnote}
decay is color-suppressed, the decay rate 
is much smaller
than those of the $\Bz\to\rho^{+}\rho^{-}$ and $B^{+}\to\rho^{+}\rho^0$
channels~\cite{rho0rho0babar,rho0rhopbelle,rho0rhop2,rhoprhomPRD,rhoprhombelle},
and is sensitive to the penguin amplitude.
Therefore a stringent limit on
$\Delta\alpha$ can be set~\cite{gronau90, rhoprhomPRD, falketal}.
This makes the $\rho\rho$ system particularly effective for
measuring~$\alpha$.

In $B\to\rho\rho$ decays the final state is
a superposition of \CP-odd and \CP-even states. 
An isospin-triangle relation~\cite{gronau90} holds for each
of the three helicity amplitudes $A_{\lambda=-1,0,+1}$, which can be
separated through an angular analysis. The longitudinal polarization
fraction $f_L=|A_0|^2/\sum|A_\lambda|^2$ can be determined through
the measurement of the distribution of the helicity angles $\theta_1$
and $\theta_2$, defined 
as the angles between the direction of the $\pi^+$ and the direction 
of the \B\ meson in the rest system of each of the $\rho^0\to\pi^+\pi^-$ candidates.

In this paper, we update our previous measurement~\cite{rho0rho0babar}
of the branching 
fraction \BR\ and  longitudinal polarization fraction $f_L$ in
$B^0\to\rho^0\rho^0$ decays, along with \BR\ for $B^0\to\rho^0 f_0$,
$f_0f_0$, $\rho^0\pi^+\pi^-$, and $\pi^+\pi^-\pi^+\pi^-$. In addition, we present the first
study of the time-dependent $C\! P$ asymmetry $\mathcal{A}_{C\!P}$ in
this mode. 
We determine the coefficients $C_L^{00}$ and $S_L^{00}$ of 
$\mathcal{A}_{C\!P}$ for the longitudinal component, expressed as a
function of $\Delta t$, the proper time difference between 
the two $B$ decays in $\Y4S\to\Bz\Bzb$:
\begin{equation}
\mathcal{A}_{C\!P}(\Delta t)=-C^{00}_L \cos{\Delta m\Delta t}+S^{00}_L\sin{\Delta m\Delta t}\ .
\label{eq:dt}
\end{equation}
where $\Delta m=(0.507\pm0.005)\hbar~\mathrm{ps}^{-1}$ is the mass
difference between two $B^0$ mass eigenstates~\cite{PDG2006}. 
When combined with measurements of 
$B^0\to\rho^+\rho^-$ and $B^+\to\rho^+\rho^0$, 
$\mathcal{A}_{C\!P}$, \BR, and
$f_L$ in $B^0\to\rho^0\rho^0$
allow a complete isospin analysis  and improve the constraints
on the penguin contribution to $B\to\rho\rho$ decays.
Changes with respect to our previous analysis~\cite{rho0rho0babar} 
include a larger data sample, improved track-selection techniques,
and inclusion of the $B$-decay time information.

%%%%%%%%%%%%%%%%%%%%%%%%%%%%%%%%%%%%%%%%%%%%%%%%%%%%%%%%%%%%%%%%%%%%%%%%%
% DETECTOR AND DATASET
%%%%%%%%%%%%%%%%%%%%%%%%%%%%%%%%%%%%%%%%%%%%%%%%%%%%%%%%%%%%%%%%%%%%%%%%%

We use a sample of $(465\pm 5)\times10^6$  \FourS\ decays into \BB\
 pairs collected with the \babar\ detector~\cite{babar} 
at the \pep2\ asymmetric-energy \epem\ collider~\cite{pep2}.
A detailed description of the \babar\ detector is available
elsewhere~\cite{babar,rho0rho0babar}. 

%%%%%%%%%%%%%%%%%%%%%%%%%%%%%%%%%%%%%%%%%%%%%%%%%%%%%%%%%%%%%%%%%%%%%%%%%
% ANALYSIS METHOD
%%%%%%%%%%%%%%%%%%%%%%%%%%%%%%%%%%%%%%%%%%%%%%%%%%%%%%%%%%%%%%%%%%%%%%%%%

We select $\B\to M_1M_2\to(\pi^+\pi^-)(\pi^+\pi^-)$
candidates, where $M_{1,2}$ stands for a $\rho^0$ or $f_0(980)$ candidate,
from neutral combinations of four charged tracks that
are consistent with originating from a single vertex near
the $e^+e^-$ interaction point. We veto tracks that are positively
identified as kaons or electrons.
The identification of signal $B$ candidates is based
on several kinematic variables. 
The beam-energy-substituted mass
$\mes = [(s/2 + {\mathbf {p}}_i\cdot {\mathbf{p}}_B)^2/E_i^2-
{\mathbf {p}}_B^2]^{1/2}$,
where the initial $e^+e^-$
four-momentum $(E_i, {\mathbf {p_i}})$ and the \B
momentum ${\mathbf {p_B}}$ are defined in the laboratory frame, is
centered near the \B mass with a resolution of $2.6~\mevcc$ for signal
candidates.  
The difference $\DeltaE = E_B^{\rm cm} - \sqrt{s}/2$ between the
reconstructed \B energy in the 
center of mass (c.m.) frame and its known value $\sqrt{s}/2$ has a maximum near zero with a
resolution of $20~\mev$ for signal events. Four other kinematic
variables describe two possible
$\pi^+\pi^-$ pairs: invariant masses $m_{1}$, $m_{2}$
and helicity angles $\theta_1,\ \theta_2$. 

We use the kinematic selection of signal candidates described
in~\cite{rho0rho0babar}. We require
$5.245 < \mes < 5.290~\gevcc$, 
$|\DeltaE|<$ 85~\mev,
$0.55< m_{1,2} < 1.050~\gevcc$,
and $|\cos\theta_{1,2}|<0.98$.
The extended di-pion invariant mass range allows us to study
the non-resonant decays $B^0\to \rho^0\pi^+\pi^-$ and
$B^0\to\pi^+\pi^-\pi^+\pi^-$, as well as $B^0\to\rho^0f_0$ 
and $B^0\to f_0f_0$. 
The contributions from the
higher mass resonances in this range are relatively small. 
We suppress the dominant  $\epem\to q\bar{q}\ (q=u,d,s,c)$ continuum
background using  
a neural network-based discriminant $\mathcal{E}$, which combines eight
topological variables~\cite{rho0rho0babar}.

We use multivariate $B$-flavor tagging
algorithms trained to identify primary leptons, kaons, soft pions,
and high-momentum charged particles
from the other $B$, called $B_{\rm tag}$~\cite{babarsin2beta}.
The effective tagging efficiency
is $(31.1\pm0.3)\%$.
Additional background discrimination power arises from the difference 
between the tagging efficiencies for signal and background in seven
tagging categories ($c_{\rm tag}$).
We determine $\Delta t$ and its error $\sigma_{\Delta t}$ 
from the spatial separation between 
the decay vertices of the signal $B$ and $B_{\rm tag}$ and require 
$|\Delta t|<15$~ps and $\sigma_{\Delta t}<2.5$~ps.

After application of all selection criteria,
72154 events are retained.
On average, each selected event has $1.05$ signal candidates, 
while in Monte Carlo (MC)~\cite{EvtGen,GEANT}
samples of longitudinally (transversely)
polarized $B^0\to\rho^0\rho^0$ decays
we find $1.15$ ($1.03$) candidates.
When more than one candidate is present in the same event,
the candidate that yields the smallest $\chi^2$ 
for the four-pion vertex is selected. 
Simulation shows that 18\% of longitudinally
and 4\% of transversely polarized $\Btozz$
events are misreconstructed with one or more tracks
that do not originate from the $B^0\to\rho^0\rho^0$ decay.
These are mostly 
low-momentum tracks from the other \B meson in the event. Such
partially reconstructed candidates are included in the definition of
the signal probability density functions (PDFs). 

%%%%%%%%%%%%%%%%%%%%%%%%%%%%%%%%%%%%%%%%%%%%%%%%%%%%%%%%%%%%%%%%%%%%%%%%%
% MAXIMUM LIKELIHOOD FIT
%%%%%%%%%%%%%%%%%%%%%%%%%%%%%%%%%%%%%%%%%%%%%%%%%%%%%%%%%%%%%%%%%%%%%%%%%

We use an unbinned extended maximum likelihood fit to extract
the $B^0\to\rho^0\rho^0$ event yield, $f_L$, $C_L^{00}$, and $S_L^{00}$.
We also fit for the event yields of
$B^0\to\rho^0f_0$, $B^0\to f_0f_0$, $B^0\to \rho^0\pi^+\pi^-$, and
$B^0\to\pi^+\pi^-\pi^+\pi^-$ decays.  
The likelihood function 
includes
the background components from non-signal $B$ decays and continuum. The 
PDFs for each component depend on
ten discriminating variables: $m_{\rm{ES}}, \Delta E, \mathcal{E},
m_1, m_2, \cos\theta_1, \cos\theta_2, c_{\rm tag},
\Delta t$, and $\sigma_{\Delta t}$.

Since the statistical correlations among the variables are found to be
negligibly small,
we take the total PDF as the product of the PDFs for the
separate variables. Exceptions are the kinematic correlation between the two
helicity angles in signal, and mass-helicity correlations in
several $B$-decay classes and misreconstructed signal. 

We use double-Gaussian functions to parameterize the
$m_{\rm{ES}}$ and $\Delta E$ PDFs for signal,
and relativistic Breit-Wigner functions 
for the resonance line-shapes of $\rho^0$
and $f_0(980)$, with the $f_0(980)$ mass and width taken from~\cite{f0mass}.
The angular distribution for 
$B$ decays~\cite{EvtGen} 
(expressed as a function of $f_L$
for \Btozz) is multiplied by a detector acceptance function 
determined from simulations. We assume that the $\rho^0$ in
$B^0\to\rho^0\pi^+\pi^-$ is longitudinally-polarized (\ie, $\pi^+\pi^-$
are produced in the S-wave), and we use the phase-space distributions
for $B^0\to\pi^+\pi^-\pi^+\pi^-$. 
The $(\pi\pi)$ invariant mass and helicity distributions of misreconstructed
signal events 
are parameterized with empirical shapes in a way similar
to that used for $B$ background discussed below.
The neural network discriminant ${\cal E}$ 
is described by the sum of three asymmetric
Gaussian functions with different parameters for signal
and background distributions.

The PDFs for  non-signal \B-decay modes are
generally modeled with empirical analytical distributions.
Several variables have distributions
identical to those for signal, such as $m_{\rm{ES}}$
when all four tracks come from the same $B$, or 
$m_{1,2}$ when both tracks come from
a $\rho^0$ meson.
Also for some of the modes the two $\pi^+\pi^-$ pairs
can have different mass and helicity distributions, 
\eg, when only one of the two combinations 
comes from a genuine $\rho^0$ or $f_0$ meson, 
or when one of the two pairs contains a
high-momentum pion (as in $B\to a_1\pi$). In such cases,
we use a four-variable correlated mass-helicity PDF.
The proper-time distribution for signal and background \B\ decays 
is convolved with a resolution function~\cite{babarsin2beta}, 
while the time distribution of continuum 
background is assumed to have zero lifetime. 

The signal and $B$-background PDF parameters are extracted from
simulation. The MC parameters for
the $m_{\rm{ES}}$, $\Delta E$, and ${\cal E}$ PDFs are adjusted by
comparing data and simulation in control channels with similar
kinematics and topology,
such as $B^0\to D^-\pi^+$ with $D^-\to K^+\pi^-\pi^-$.
The continuum background PDF shapes are extracted 
from on-resonance sideband data ($\mes<5.27\gevcc$), with parameters
of most PDFs (for $m_\mathrm{ES}$, $\Delta E$, ${\cal
  E}$, $\theta_1$, $\theta_2$, and $\Delta t$) left free in the final fit. 
The tagging efficiencies, mistag fractions, and the parameters of the
proper-time 
distributions for signal modes are 
obtained in dedicated fits to events with identified exclusive \B\
decays~\cite{babarsin2beta}. 
For inclusive \B\ backgrounds these parameters are determined 
by MC and their systematic uncertainties are evaluated in data.

%%%%%%%%Fitter Description%%%%%%%%
We study the contributions of the dominant backgrounds by using 
high-statistics exclusive MC samples. We single out the $B^0\to
a_1^{\pm}\pi^{\mp}$,  
$B^0\to \rho^0K^{*0}$
and $B^0\to f_0K^{*0}$ modes, which have kinematic distributions similar to
those of the signal events, and parameterize their PDFs individually. The event
yield for $B^0\to a_1^{\pm}\pi^{\mp}$
is allowed to vary in the fit, while the yields for 
$B^0\to \rho^0K^{*0}$, $B^0\to f_0K^{*0}$ are fixed to the expected values 
~\cite{kstrho}.

In addition, we construct a charmless event category consisting of 
$B^0\to\rho^+\rho^-$, $B^0\to \rho^{\pm}\pi^{\mp}$, $B^+\to\rho^+\rho^0$, 
$B^+\to a_1^{0}\pi^{+}$, $B^+\to a_1^{+}f_0$, $B\to\eta'K$, and
$B^+\to\rho^0\pi^+$ backgrounds. Kinematic distributions in these events, 
especially events in which at
least one charged particle is not correctly associated to the \B
candidate, are similar to each other, and
also to those of other, poorly measured, charmless decays. We allow
the overall event yield for this category of events to vary in the
fit, after fixing 
the relative weight of each mode to the expected
value~\cite{PDG2006, HFAG07}. The remaining events, which mostly originate
from open $b\to c$ transitions, are parameterized as a
separate background component, with its yield left free in the
fit. 

\begin{table*}[ht!]
\caption{
Event yields;
fraction of longitudinal polarization ($f_L$);
selection efficiency (Eff) corresponding to measured polarization;
branching fractions (${\cal B}$);
branching fraction upper limits (UL) at 90\% confidence level (CL);
and significance $\mathcal{S}$ including systematic uncertainties.
First uncertainty is statistical and second is systematic. 
}
\begin{center}
\begin{tabular*}{\textwidth}{@{\extracolsep{\fill}}lc c c c c c}
\hline\hline
Mode\rule{0pt}{12pt} & Yield & $f_L$ & Eff (\%) & 
$\mathcal{B}~(10^{-6})$ & UL $(10^{-6})$ & $\mathcal{S}~(\sigma)$ \\
\hline
&&&&&&\vspace*{-3mm}\\
$B^0\to\rho^0\rho^0$ & $99^{+35}_{-34}\pm 15$ & $0.75^{+0.11}_{-0.14}\pm 0.04$  & $23.28\pm 0.07$ &
$0.92\pm0.32\pm 0.14$  &-& $3.1$ \\[0.5mm]
$B^0\to\rho^0f_0\to\rho^0[\pi^+\pi^-]_{f_0}$ & $3^{+22}_{-20}\pm 5$ &-& $24.16\pm 0.09$ & 
$0.03^{+0.20}_{-0.18}\pm 0.05$ & $<0.34$ &  \\[0.5mm]
$B^0\to f_0f_0\to[\pi^+\pi^-]_{f_0}[\pi^+\pi^-]_{f_0}$ & $6^{+8}_{-5}\pm 2$ &-& $27.22\pm 0.07$ & 
$0.05^{+0.06}_{-0.04}\pm 0.02$ & $<0.16$ &  \\[0.5mm]
$B^0\to\rho^0\pi^+\pi^-$ & $-12^{+39}_{-35}\pm 9$ &-& $1.68\pm 0.01$ & 
$-1.2^{+5.0}_{-4.5}\pm 1.1$ & $<8.7$ &  \\[0.5mm]
$B^0\to\pi^+\pi^-\pi^+\pi^-$ & $8^{+30}_{-25}\pm 6$ &-& $0.55\pm 0.01$ & 
$3.2^{+11.7}_{-9.8}\pm 3.4$ & $<21.1$ &  \\[0.5mm]
\hline\hline
\end{tabular*}
  \label{tab:results}
\vspace{-0.5cm}
\end{center}
\end{table*}
%%%%%%%%%%%%%%%%%%%%%%%%%

%%%%%%%%%%%%%%%%%%%%%%%%%%%%%%%%%%%%%%%%%%%%%%%
\begin{figure}[tb]
\begin{center}
\setlength{\epsfxsize}{\linewidth}\leavevmode\epsfbox{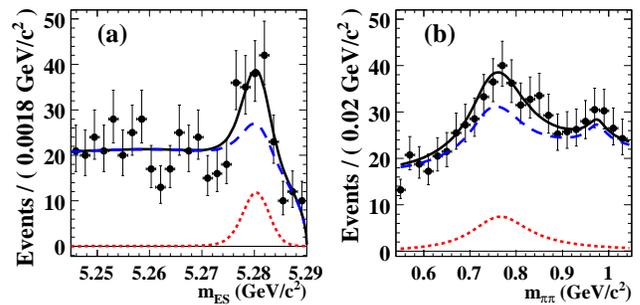}
\end{center}
\vspace{-0.5cm}
\caption{
Projections of the multidimensional fit onto the
(a) $m_{\rm ES}$, and
(b) di-pion invariant mass $m_{\pi\pi}$ (average of $m_1$ and $m_2$
distributions is shown),
after a requirement on the 
signal-to-background probability ratio
with the plotted variable excluded,
which enhances the fraction of
signal events in the sample. This selection has 39\% (60\%) efficiency for
signal for the $m_{\rm ES}$ ($m_{\pi\pi}$) projection. 
The data points are overlaid by the full PDF projection (solid black
line). Also shown are the  $B^0\to\rho^0\rho^0$ PDF component
(dotted line) and the sum of all other PDFs (dashed line). 
}
\vspace{-0.5cm}
\label{fig:projections}
\end{figure}
%%%%%%%%%%%%%%%%%%%%%%%%%%%%%%%%%%%%%%%%%%%%%%%

%%%%%%%%%%%%%%%%%%%%%%%%%%%%%%%%%%%%%%%%%%%%%%%%%%%%%%%%%%%%%%%%%%%%%%%%%
% PHYSICS RESULTS
%%%%%%%%%%%%%%%%%%%%%%%%%%%%%%%%%%%%%%%%%%%%%%%%%%%%%%%%%%%%%%%%%%%%%%%%%

Table~\ref{tab:results} summarizes the results of the fit.
The $\Bztorhozrhoz$ decay is observed with a significance of $3.1$
standard deviations ($\sigma$),
as determined by 
the quantity $\mathcal{S} =
\sqrt{-2\ln(\mathcal{L}_0/\mathcal{L}_{\max})}$, where  
$\mathcal{L}_{\max}$ is the maximum likelihood value, and 
$\mathcal{L}_0$ is the likelihood for a fit with the signal
contribution set to zero. Both likelihoods include systematic
uncertainties, which are assumed to be Gaussian-distributed and are discussed below. 
This significance level corresponds to a probability of
$1.0\times10^{-3}$ that the observed signal yield is consistent with a
background fluctuation. We do not
observe significant event yields for $\Bztorhozfz$ or $\Bztofzfz$
decays, nor for the non-resonant decays $B^0\to \rho^0\pi^+\pi^-$ and
$B^0\to\pi^+\pi^-\pi^+\pi^-$. 
We find $280\pm53$,
$670\pm96$, $2329\pm147$, and $68701\pm281$ events for the
$B^0\to a_1^{\pm}\pi^{\mp}$, charmless, $b\to c$, and continuum
backgrounds, respectively, consistent with expectations. 
In Fig.~\ref{fig:projections} we show the projections of the fit
results onto $m_{\rm ES}$ and $m_{\pi\pi}$.

From the fit to the proper-time distribution of the data sample, we
determine the $C\! P$-violating parameters 
\begin{eqnarray}
S^{00}_L &=& 0.3\pm 0.7\ (\mathrm{stat.})\pm 0.2\ (\mathrm{syst.}) \nonumber \\
C^{00}_L &=& 0.2\pm 0.8\ (\mathrm{stat.})\pm 0.3\ (\mathrm{syst.}) \nonumber 
\end{eqnarray}
for the longitudinal component of the $\B^0\to\rho^0\rho^0$
sample. 
%
% correlations
%
\input corr

%%%%%%%%%%%%%%%%%%%%%%%%%%%%%%%%%%%%%%%%%%%%%%%%%%%%%%%%%%%%%%%%%%%%%%%%%
% SYSTEMATIC STUDIES
%%%%%%%%%%%%%%%%%%%%%%%%%%%%%%%%%%%%%%%%%%%%%%%%%%%%%%%%%%%%%%%%%%%%%%%%%

Dominant systematic uncertainties in the fit originate from 
statistical errors in the PDF parameterizations due to the limited
number of events in the control samples, variations in the \B
background branching ratios fixed in the fit, and from potential fit bias. 
The PDF parameters are varied by their respective uncertainties
to derive the corresponding systematic errors. 
The fit bias is studied in a large number ($\sim 1000$) of MC experiments,
in which signal and dominant charmless \B background events are fully
simulated, while other backgrounds are sampled from their respective PDFs. 
We correct for the bias of $7.9\pm 2.0$ events for $B^0\to\rho^0\rho^0$, $0.03\pm 0.02$ 
for $C^{00}_L$ and $0.07\pm 0.03$ for $S^{00}_L$. 
The uncertainties associated with the \B background model
are $4$ events for the signal yield, $0.01$ for $f_L$, $0.01$ for
$C^{00}_L$ and $0.11$ for $S^{00}_L$. 
The systematic uncertainties due to the charmless background
composition, arising from the uncertainties in the individual
branching ratios and the $C\! P$ content of the \B
background~\cite{HFAG07,ref:a1piCP}, are  
$5$ events for the signal yield, $0.01$ for $f_L$, $0.18$ for
$C^{00}_L$ and $0.14$ for $S^{00}_L$.
The above systematic uncertainties do not scale with event yield
and are included in the calculation of the significance of the result.

We estimate the systematic uncertainty due to the interference 
between the $B^0\to\rho^0\rho^0$ and $a_1^{\pm}\pi^{\mp}$ decays using 
simulated samples in which the decay amplitudes for $B^0\to\rho^0\rho^0$
are generated according to this measurement
and those for $B^0\to a_1^{\pm}\pi^{\mp}$ correspond
to a branching fraction of $(33.2\pm4.8)\times 10^{-6}$~\cite{a1pi}.
The strong phases and \CP\ content of the interfering 
$a_1^{\pm}\pi^{\mp}$ state are varied between zero and a maximum 
value using uniform distributions.
We take the RMS variation of the fitted values
(13 events for the $\rho^0\rho^0$ yield, $0.02$ for $f_L$, and $0.04$
for $S_L^{00}$ and $C_L^{00}$) 
as a systematic uncertainty.

Uncertainties in the reconstruction efficiency
arise from track finding and particle identification, and are
determined by dedicated studies on control data
samples. Uncertainties due to other selection requirements,
such as vertex probability, track multiplicity, and thrust angle,
amount to $2.4\%$ for the event yields, and are negligible for the
polarization and $C\! P$ observables. 

%%%%%%%%%%%%%%%%%%%%%%%%%%%%%%%%%%%%%%%%%%%%%%%%%%%%%%%%%%%%%%%%%%%%%%%%%
% ALPHA
%%%%%%%%%%%%%%%%%%%%%%%%%%%%%%%%%%%%%%%%%%%%%%%%%%%%%%%%%%%%%%%%%%%%%%%%%

We perform an isospin analysis of  $B\to\rho\rho$ decays,
by minimizing a $\chi^2$ that includes the measured quantities
expressed as the lengths of the sides of the isospin triangles~\cite{gronau90}.
We use the measured branching fraction and
fraction of longitudinal polarization of 
$\Bptorhozrrhop$ decays~\cite{rho0rhop2},  
the measured branching fraction, polarization, and 
 \CP\ parameters $S^{+-}_{L}$ and $C^{+-}_{L}$
determined from the time evolution of the longitudinally
polarized $\Bztorhoprhom$ decays~\cite{rhoprhomPRD}, 
and the branching fraction, polarization,
and \CP\ parameters $S^{00}_{L}$ and $C^{00}_{L}$
of $\Bztorhozrhoz$ reported here. 
We assume uncertainties to be Gaussian distributed and 
neglect $I=1$ isospin contributions, electroweak loop amplitudes, 
non-resonant, and isospin-breaking effects.

\begin{figure}[tb!]
\begin{center}
\setlength{\epsfxsize}{0.95\linewidth}
\leavevmode\epsfbox{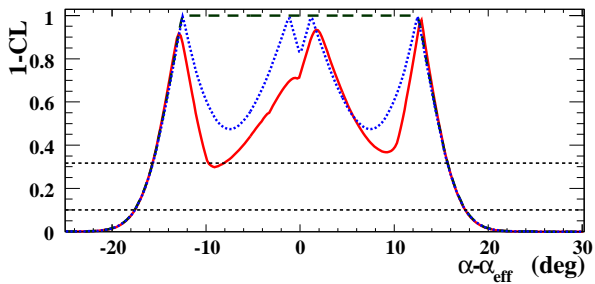}
\vspace{-0.3cm}
\caption{Confidence level (CL) on $\alpha-\alpha_{\rm eff}$ determined from 
the isospin analysis. 
The long-dashed curve is obtained without 
the two \CP\ parameters 
$S^{00}_{L}$ and $C^{00}_{L}$. 
The dotted curve corresponds to the isospin analysis without
$S^{00}_L$, and the solid curve CL
includes both $C^{00}_L$ and $S^{00}_L$ in the fit. 
The horizontal dotted lines correspond to the $68\%$ (top) and $90\%$ 
(bottom) CL  intervals.}
\label{fig:DeltaAlphaScan}
\vspace{-0.5cm}
\end{center}
\end{figure}

We obtain a 68\% (90\%) CL limit
$|\alpha-\alpha_{\rm eff}|<15.7^\circ$ ($<17.6^\circ$)
where $\alpha_{\rm eff}$ is defined by
$\sin(2\alpha_{\rm eff})= S^{+-}_{L}/({1-C^{+-2}_{L}})^{1/2}$.
Fig.~\ref{fig:DeltaAlphaScan} shows the confidence level with 
and without use of $S^{00}_{L}$ and $C^{00}_{L}$
in the isospin analysis fit. We observe four solutions near zero,
as in the $B\to\pi\pi$ isospin analysis~\cite{pi0pi0}.
The  additional constraint from $S^{00}_L$ 
provides some discrimination among the four solutions. 

%%%%%%%%%%%%%%%%%%%%%%%%%%%%%%%%%%%%%%
%%%%%%%%%%%%%%%%%%%%%%%%%%%%%%%%%%%%%%

%%%%%%%%%%%%%%%%%%%%%%%%%%%%%%%%%%%%%%%%%%%%%%%%%%%%%%%%%%%%%%%%%%%%%%%%%
% SUMMARY
%%%%%%%%%%%%%%%%%%%%%%%%%%%%%%%%%%%%%%%%%%%%%%%%%%%%%%%%%%%%%%%%%%%%%%%%%

In summary, we confirm our earlier evidence ~\cite{rho0rho0babar} for 
\Btozz\ decays with a significance of $3.1\sigma$ 
and measure the branching fraction, longitudinal polarization
fraction, and $C\! P$ asymmetries in these decays. 
These measurements combined with those for
$B^+\to\rho^0\rho^+$ and $B^0\to\rho^+\rho^-$ decays provide
a constraint on the penguin uncertainty
in the determination of the CKM unitarity angle $\alpha$.
We find no significant evidence for the decays $B^0\to \rho^0f_0$,
$B^0\to f_0f_0$, $B^0\to \rho^0\pi^+\pi^-$ or
$B^0\to\pi^+\pi^-\pi^+\pi^-$.

%%%%%%%%%%%%%%%%%%%%%%%%%%%%%%%%%%%%%%%%%%%%%%%%%%%%%%%%%%%%%%%%%%%%%%%%%
% ACKNOWLEDGMENTS
%%%%%%%%%%%%%%%%%%%%%%%%%%%%%%%%%%%%%%%%%%%%%%%%%%%%%%%%%%%%%%%%%%%%%%%%%

We are grateful for the excellent luminosity and machine conditions
provided by our \pep2\ colleagues,
and for the substantial dedicated effort from
the computing organizations that support \babar.
The collaborating institutions wish to thank
SLAC for its support and kind hospitality.
This work is supported by
DOE
and NSF (USA),
NSERC (Canada),
IHEP (China),
CEA and
CNRS-IN2P3
(France),
BMBF and DFG
(Germany),
INFN (Italy),
FOM (The Netherlands),
NFR (Norway),
MIST (Russia),
MEC (Spain), and
PPARC (United Kingdom).
Individuals have received support from the
Marie Curie EIF (European Union) and
the A.~P.~Sloan Foundation.

%%%%%%%%%%%%%%%%%%%%%%%%%%%%%%%%%%%%%%%%%%%%%%%%%%%%%%%%%%%%%%%%%%%%%%%%%
% BIBLIOGRAPHY
%%%%%%%%%%%%%%%%%%%%%%%%%%%%%%%%%%%%%%%%%%%%%%%%%%%%%%%%%%%%%%%%%%%%%%%%%

\onecolumngrid 
\end{document}

%% file: authors_jul2008.tex
%% author list as of 02-Jul-2008 (523 authors)
%
\author{B.~Aubert}
\author{M.~Bona}
\author{Y.~Karyotakis}
\author{J.~P.~Lees}
\author{V.~Poireau}
\author{E.~Prencipe}
\author{X.~Prudent}
\author{V.~Tisserand}
\affiliation{Laboratoire de Physique des Particules, IN2P3/CNRS et Universit\'e de Savoie, F-74941 Annecy-Le-Vieux, France }
\author{J.~Garra~Tico}
\author{E.~Grauges}
\affiliation{Universitat de Barcelona, Facultat de Fisica, Departament ECM, E-08028 Barcelona, Spain }
% ...
\collaboration{The \babar\ Collaboration}
\noaffiliation

%% file: corr.tex
The statistical correlations between fit parameters for
$\B^0\to\rho^0\rho^0$ are given in Table~\ref{tab:corr}.
\begin{table}[ht]
\caption{Correlation matrix for $B^0\to\rho^0\rho^0$
  parameters.}
\begin{center}
\begin{tabular}{c|rrrr}
\hline
   Parameter
 & Yield
 & $f_L$
 & $S_L^{00}$
 & $C_L^{00}$ \cr
\hline
Yield     & $1.000$  & $0.086$  & $-0.136$ & $-0.273$ \\
$f_L$     &          & $1.000$  & $-0.006$ & $-0.174$ \\
$S_L^{00}$           &          &          & $1.000$  & $-0.035$ \\
$C_L^{00}$           &          &          &          & $1.000$  \\
\hline
\end{tabular}
\end{center}
\label{tab:corr}
\end{table}

%% file: paper_prl.bbl
\begin{thebibliography}{99}

\bibitem{CabibboKobayashi}
N. Cabibbo, Phys. Rev. Lett. {\bf 10}, 531 (1963);
M. Kobayashi and T. Maskawa, Prog. Theor. Phys. {\bf 49}, 652 (1973).

\bibitem{gronau90}
M.~Gronau and D.~London, \jprl{65}, 3381 (1990).

\bibitem{footnote}
Charge conjugate decay modes are implied in this paper.

\bibitem{rho0rho0babar}
\babar\ Collaboration, B.~Aubert {\it et al.},
\jprl{98}, 111801 (2007).

\bibitem{rho0rhopbelle}
Belle Collaboration, J.~Zhang {\it et al.},
\jprl{91}, 221801 (2003).

\bibitem{rho0rhop2}
\label{ref:rho0rhop2}
\babar\ Collaboration, B.~Aubert {\it et al.},
\jprl{97}, 261801 (2006).

\bibitem{rhoprhomPRD}
\babar\ Collaboration, B.~Aubert {\it et al.},
Phys. Rev. D {\bf 76}, 052007 (2007). 

\bibitem{rhoprhombelle}
\label{ref:rhoprhombelle}
Belle Collaboration, A.~Somov {\it et al.},
\jprl{96}, 171801 (2006).

\bibitem{falketal}
A.F.~Falk, Z.~Ligeti, Y.~Nir, H.~Quinn, Phys.\ Rev.\ D {\bf 69}, 011502(R) (2004).

\bibitem{PDG2006}
Particle Data Group,  W.-M. Yao {\it et al.}, J. Phys. G {\bf 33}, 1 (2006).

\bibitem{babar}
\babar\ Collaboration, B.~Aubert {\it et al.},
Nucl. Instrum. Methods Phys. Res.,
Sect. A {\bf 479}, 1 (2002).

\bibitem{pep2}
PEP-II Conceptual Design Report, SLAC-R-418 (1993).

\bibitem{babarsin2beta}
\babar\ Collaboration, B.~Aubert {\it et al.},
\jprl{94}, 161803 (2005).

\bibitem{EvtGen}
D.\ Lange, Nucl. Instrum. Methods Phys. Res.,
Sect. A {\bf 462}, 152 (2001).


\bibitem{GEANT}
The \babar\ detector Monte Carlo simulation is based
on GEANT4: S. Agostinelli {\it et al.},
Nucl. Instrum. Methods Phys. Res.,
Sect. A {\bf 506}, 250 (2003).

\bibitem{f0mass}
E791 Collaboration, E. M. Aitala {\it et al.},
Phys. Rev. Lett. {\bf 86}, 765 (2001).

\bibitem{kstrho}
\babar\ Collaboration, B.~Aubert {\it et al.},
\jprl{97}, 201801 (2006).

\bibitem{HFAG07}
Heavy Flavor Averaging Group,
E. Barberio {\it et al.}, arXiv:0704.3575 [hep-ex] and online update 
at http://www.slac.stanford.edu/xorg/hfag .

\bibitem{ref:a1piCP}
\babar\ Collaboration, B.~Aubert {\it et al.},
\jprl{98}, 181803 (2007).

\bibitem{a1pi}
\babar\ Collaboration, B.~Aubert {\it et al.},
\jprl{97}, 051802 (2006).

\bibitem{pi0pi0}
\babar\ Collaboration, B.~Aubert {\it et al.},
Phys. Rev. D {\bf 76}, 091102 (2007).

\end{thebibliography}
